# Cross-country structure of a sexual contact network derived from a commercial-sex review platform

Hiromu Ito[1,2*], Keiko Shigeta[1], Toshiaki R. Asakura[3,4,5], Satoru Morita[6*]

[1]Department of International Health and Medical Anthropology, Institute of Tropical Medicine, Nagasaki University, Nagasaki, Japan

[2]Multidisciplinary Pandemic Research Center, Nagasaki University, Nagasaki, Japan

[3]Department of Infectious Disease Epidemiology and Dynamics, London School of Hygiene & Tropical Medicine, London, UK

[4]Centre for Mathematical Modelling of Infectious Diseases, London School of Hygiene & Tropical Medicine, London, UK

[5]School of Tropical Medicine and Global Health, Nagasaki University, Nagasaki, Japan

[6]Department of Mathematical and Systems Engineering, Shizuoka University, Shizuoka, Japan

* Corresponding author*Hiromu Ito, *Satoru Morita

**Email:** ito.hiromu@nagasaki-u.ac.jp (HI), morita.satoru@shizuoka.ac.jp (SM)

**Abstract**

Sexual contact networks are fundamental to understanding the spread of sexually transmitted infections (STIs), yet their large-scale structure remains difficult to observe because of technical, ethical, and privacy constraints. Here, we analyze a large bipartite commercial-sex contact network comprising 1,436,738 client–sex worker links, constructed from reviews posted between 1999 and 2024 on *The Erotic Review* (*TER*), a review platform covering sex workers listed across 97 countries and territories. We examine the network's cross-country structure, focusing on how clients, sex workers, locations, and time periods are connected. The network formed a giant connected component containing most clients and sex workers and exhibited small-world-like properties, including short average path lengths, high bipartite clustering, and disassortative degree correlations. Although most reviews were associated with the United States, a small fraction of clients reviewed sex workers in multiple countries or territories, creating spatial bridges between otherwise distant regions. Long-term users created temporal bridges, making the aggregated network more compact than networks built from shorter time windows. Contacts involving transgender female sex workers were structurally concentrated: clients who reviewed both cisgender and transgender female sex workers represented a small minority but accounted for a large share of transgender-related reviews and showed higher cross-country activity. These findings show how a small number of highly connected clients with cross-country review activity can shape the large-scale connectivity of commercial-sex contact networks. More broadly, our results demonstrate the value and limitations of digital trace data for studying cross-country contact structures relevant to STI transmission.

## Introduction

The structure of sexual contact networks is a key determinant of how sexually transmitted infections (STIs) spread through a population, since epidemic dynamics on a network depend not only on individual risk behavior but on how individuals are connected to one another [1,2]. Contact tracing following STI outbreaks has repeatedly shown that diagnosed individuals are often embedded in large, interconnected clusters linked through chains of sexual contact [3–11], underscoring that sexual contact should be understood as a network phenomenon rather than a set of isolated events. Characterizing the large-scale structure of these networks is therefore essential for public health modeling and intervention design.

Progress on this front has been constrained by the difficulty of observing sexual contact networks at scale. Partner-notification data are limited to individuals already diagnosed with an STI and capture only local, outbreak-specific fragments of the network [3–11]. Population-level surveys, such as those conducted in Sweden [12], the UK, Zimbabwe [13], Burkina Faso [14], and Japan [15], instead ask respondents to report their number of partners without identifying them, and consistently find that partner counts follow a heavy-tailed distribution—a small fraction of individuals report disproportionately many partners, a pattern known to facilitate epidemic spread even under low transmission probabilities [1,2]. However, because these surveys record degree but not ties between specific individuals, they provide no information about the network's topology: connectivity, path lengths, clustering, or the existence of bridges between otherwise separate parts of the population remain unobserved.

Online review platforms, on which clients pseudonymously report their encounters with sex workers, offer a way to construct explicit sexual contact networks at scale [16]. Studies using such data have built bipartite networks with clients and sex workers comprising thousands to tens of thousands of individuals in Brazil [17], the UK [18], and Japan [19], as well as MSM commercial sex networks in the U.S. [20]. A consistent finding across these studies is that sex workers form localized clusters, while a subset of clients bridges otherwise disconnected parts of the network by engaging with sex workers across multiple locations [17,19,20], contributing disproportionately to overall connectivity and outbreak potential. However, each of these networks was constructed within a single country or metropolitan area. Whether and how such bridging extends across national borders, linking sex workers and clients across countries into a single large-scale network, has not been examined, leaving the cross-country structure of sexual contact networks unresolved.

We address this gap using data from *The Erotic Review* (*TER*), an online review platform active since 1999. Although *TER* data are heavily centered on the U.S., the platform also contains substantial review activity involving sex workers listed in many other countries and territories. *TER* has previously supported research on the sex industry's digital transformation and client behavior [21–27], but its international coverage has not yet been used to examine the cross-country topology of a single sexual contact network.

In this study, we construct a large bipartite sexual contact network comprising 1,436,738 client–sex worker links (derived from 1,468,852 *TER* reviews posted between April 1999 and June 2024, after removing duplicate reviews between the same pair), linking clients and sex workers across 97 countries and territories. We examine the network's cross-country structure and show that a small fraction of clients who review sex workers in multiple countries act as spatial bridges connecting otherwise distant regions, while long-term users active over many years act as temporal bridges that make the quarter-century aggregated network more compact than networks built from shorter time windows. We further show that contacts involving transgender female sex workers (TransFSW) are structurally concentrated among a small number of highly active clients with cross-country review activity. Together, these findings characterize, for the first time, the large-scale cross-country structure of a commercial sexual contact network and its implications for STI transmission across connected regions.

## Data and Methods

### *Study website*

*The Erotic Review (TER)* (https://www.theeroticreview.com/) is an online commercial-sex review platform on which clients post pseudonymous reviews of encounters with sex workers, referred to as "providers" on the website. *TER* has been active since 1999 and contains profiles of sex workers, review dates, reviewer usernames, service locations, and selected profile attributes. The platform is heavily centered on the United States but also includes profiles and reviews from many other countries and territories.

We used only information publicly available without paid membership. The publicly visible data include sex worker profiles, service locations, reviewer usernames, and the month and year of each review; we did not access the full text of reviews available to paid members. Clients must register a username to post reviews, and the website moderates reviews according to its own guidelines. Email addresses and other personally identifying information of reviewers are not visible to ordinary site visitors.

Each *TER* profile contains a field labeled "Transsexual" using the website's original terminology, for which the available responses are "No," "Yes (Preop)," "Yes (Postop)," and "I don't know." In this study, we used this field as an operational profile-based classification: profiles listed as "No" were classified as cisgender female sex workers (CisFSW), those listed as "Yes (Preop)" or "Yes (Postop)" were classified as transgender female sex workers (TransFSW), and those listed as "I don't know" were treated as an unknown category (DK). We note that this information is self-reported for the advertisement purpose on the platform. We therefore used these categories only as *TER* profile-based classifications and did not independently verify the gender identity of the individuals behind the profiles.

### *Target services*

*TER* includes several service categories, including escort, massage, and S&M. Because 91% of reviewed sex workers in our dataset listed services including “Escort,” and because previous studies of online commercial-sex networks have used escort reviews as proxies for sexual contacts [17,18], we restricted our analysis to sex workers whose listed services included “Escort.” We did not distinguish among more specific combinations of listed services, such as “Escort” and “Escort/Massage.”

### *Data collection*

Data were collected using Python 3.9.6 between 22 June and 11 July 2024. We first collected the URLs assigned to sex worker profiles on *TER* and then extracted publicly visible information from each profile and its associated review list. The collected variables included: (1) sex worker ID, (2) agency type, (3) service city and country/territory, (4) the *TER* profile category related to transgender status, (5) review date, recorded by month and year, and (6) reviewer username.

We included reviews posted up to the end of June 2024 and excluded reviews added thereafter. Because sex worker profiles on *TER* are created through reviews, we included only sex workers with at least one review at the time of data extraction. Usernames of clients and sex workers were replaced with arbitrary numerical identifiers before analysis to ensure the anonymity.

### *Network construction and analysis*

We constructed an undirected bipartite network, in which one node set consisted of clients and the other of sex workers, from the *TER* data. An edge was placed between a client and a sex worker if the client posted at least one review for that sex worker. Multiple reviews between the same client–sex worker pair were treated as a single edge, and all the analyses were conducted on an unweighted bipartite network. The degree of a node was defined as the number of distinct neighbors in this bipartite network.

To examine and visualize cross-country connectivity, we aggregated sex worker profiles by their listed country or territory and constructed a client-by-country bipartite network. We then projected this network onto the country/territory layer. In the projected network, countries and territories constituted the nodes, and an edge was placed between two countries if at least one client had review ties to sex worker profiles listed in both locations. This projection was used only to characterize cross-country connectivity between locations; the main network statistics were calculated on the individual-level bipartite network.

We analyzed both the full aggregated network and temporal subnetworks. Temporal subnetworks were constructed using reviews posted within three-year windows: 2000–2002, 2003–2005, 2006–2008, 2009–2011, 2012–2014, 2015–2017, 2018–2020, and 2021–2023. These subnetworks were used to compare the compactness and clustering of networks observed over shorter time periods with those of the long-term aggregated network.

We stratified clients according to the *TER* profile categories of the sex workers they reviewed. We first classified sex workers into three profile-based categories: CisFSW, TransFSW, and "Don't know" (DK) , as described above. We then grouped clients accordingly to whether they had reviewed at least one CisFSW profile and/or at least one TransFSW profile, regardless of whether they had also reviewed DK profiles. Clients who reviewed at least one CisFSW profile but no TransFSW profile were classified as CisFSW-only male clients (abbreviated as MC-CisFSW), those who reviewed at least one TransFSW profiles but no CisFSW profiles were classified as TransFSW-only male client (MC-TransFSW), and those who reviewed both CisFSW and TransFSW were classified as profilesCis–TransFSW male clients (MC-CisTransFSW). This classification was made regardless of any DK profiles they had reviewed. Clients who only reviewed DK sex workers were classified separately as DK-only clients. Contacts with DK sex workers were included in the overall network statistics but excluded from the main CisFSW/TransFSW comparisons.

We quantified network structure using standard measures for bipartite networks. We first derived the degree distributions by country, review patterns of clients and gender identity of sex workers. Clustering was measured using the bipartite clustering coefficient proposed by Robins and Alexander [28], defined as the proportion of three-paths involving two clients and two sex workers that are closed to form four-cycles. Degree correlation was measured as a bipartite degree correlation coefficient, defined as the Pearson correlation coefficient between the degrees of clients and sex workers connected by edges [29,30]. Because this coefficient can be influenced by the degree sequences of the two node sets, we compared the observed values with those obtained from degree-preserving randomized bipartite networks.

We generated degree-preserving randomized bipartite networks as null models for comparison with the observed networks. For each observed network or subnetwork, we created 100 randomized networks by rewiring edges while preserving the degree of every node in that network or subnetwork. We then compared the observed clustering coefficients and degree correlations with the corresponding null distributions. To further characterize degree mixing patterns, we also calculated the average nearest-neighbor degree for nodes of degree k, separately for clients and sex workers.

To assess small-world-like structure, we calculated the average shortest path length on the giant connected component of each network or temporal subnetwork. We interpreted the co-occurrence of the shorter path lengths and higher bipartite clustering values in the observed networks, when compared to degree-preserving randomized networks, as the evidence of small-world-like organization.

***Ethics and privacy***

This study analyzed network structure using publicly available information from *TER*, an online platform containing pseudonymous reviews of reported commercial-sex encounters. We used only information accessible without paid membership, including sex worker profile information, service locations, reviewer usernames, and review dates. We did not access the full text of paid reviews or any non-public information such as email addresses.

The data were handled in a de-identified form before the analysis. Client usernames and sex worker identifiers were replaced with arbitrary numerical codes, and no legal names, contact details, or other directly identifying information were included in the analytical dataset. Results are reported only in an aggregate form, and no individual client or sex worker can be identified in the manuscript.

Although the study analyzed human behavioral data, it did not involve direct participation, intervention, interaction with individuals, or collection of identifiable private information. The Human Research Ethics Committee of Shizuoka University determined that formal ethics approval and informed consent were not required because the study used de-identified, publicly available online data and did not fall under institutional definitions of human participants research requiring review. This determination was made on 21 May 2024.

## Results

We collected 1,468,852 review records connecting 254,410 clients and 275,855 sex workers. Because some client–sex worker pairs were associated with multiple reviews, we collapsed repeated reviews between the same pair into a single edge. This resulted in 1,436,738 unique client–sex worker links (Table 1). In total, 27,340 client–sex worker pairs had repeated reviews, corresponding to 32,114 duplicate review records beyond the first review for each pair. Because repeated reviews accounted for only a small fraction of all review records, we did not incorporate edge weights and instead analyzed the network as an unweighted bipartite graph.

The giant connected component (GCC) contained 244,359 clients and 264,806 sex workers, corresponding to 509,165 individuals, or 96.0% of all nodes in the review network. This proportion is substantially higher than those reported in previous studies of online commercial-sex contact networks (Table 1). The second-largest connected component contained only 12 nodes, indicating that disconnected components were extremely small relative to the GCC and highlighting the strong connectivity of the *TER*-derived network.

The listed working locations of reviewed sex workers spanned 97 countries and territories, including all 50 U.S. states and Washington, D.C. (Fig. 1). The distribution was highly concentrated in the United States, where 89.9% of reviewed sex workers were listed. Canada contained more than 10,000 reviewed sex workers, while the UK, Mexico, and Germany each contained more than 1,000.

[Figure 1 near here]

To examine cross-country connectivity, we constructed a country-level projection in which two countries or territories were connected if at least one client reviewed sex workers listed in both locations. This projection should be interpreted as a network of cross-country review ties rather than a direct record of physical travel. The resulting country-level network contained 97 countries and territories, of which 93 belonged to a single connected component. Only four locations—Jersey, Jordan, Sudan, and Trinidad and Tobago—were isolated from the main component (Fig. 2). These results indicate that, despite the strong concentration of reviews in the United States, *TER*-derived review ties connect a broad set of countries into a largely connected cross-country structure.

[Figure 2 near here]

Figure 3A shows the monthly numbers of review records, active reviewers, and reviewed sex workers. A sharp decline in review activity was observed during 2018–2019, coinciding with the temporary blocking of *TER* access from within the United States following the enactment of the Allow States and Victims to Fight Online Sex Trafficking Act (FOSTA) and the Stop Enabling Sex Traffickers Act (SESTA) in April 2018 [31]. *TER* restored access for users in the United States in December 2019 [32]. Consistent with the overall geographic distribution of the dataset, most monthly reviewers and reviewed sex workers were associated with the United States throughout the observation period (Fig. 3B,C).

[Figure 3 near here]

We next compared clients with cross-country review activity to those whose reviews were limited to sex workers listed in a single country or territory. Here, a client's degree denotes the number of distinct sex workers reviewed by that client. Among clients who posted two or more reviews, 14,199 had reviewed sex workers listed in multiple countries or territories. These clients exhibited a heavier-tailed degree distribution than single-country clients, indicating that cross-country review activity was associated with a larger number of distinct reviewed sex workers and substantial heterogeneity in client activity (Fig. 4A).

[Figure 4 near here]

We classified sex workers into three categories based on the "Transsexual" field in their public *TER* profiles (Fig. 5 and Table S1). The majority were classified as cisgender female sex workers (CisFSW; 94.8%), followed by transgender female sex workers (TransFSW; 4.3%) and an unknown category ("Don't know"; DK; 0.9%). The mean degree of sex workers, defined as the number of distinct clients who reviewed them, was 5.21 overall, 5.28 for CisFSW, 4.16 for TransFSW, and 3.09 for DK. Among TransFSW, 98% were listed as "Yes (Preop)" and 2% as "Yes (Postop)" in the *TER* profile field; their mean degrees were 4.16 and 4.20, respectively.

[Figure 5 near here]

The overall mean degree of clients was 5.65 (Table 1). We then classified clients according to the *TER* profile categories of the sex workers they reviewed (Fig. 5 and Table S2).

Most clients reviewed only CisFSW (MC-CisFSW; 93.2%), with a mean degree of 5.55. A smaller group reviewed both CisFSW and TransFSW (MC-CisTransFSW), and this group had a substantially higher mean degree of 13.6. Clients who reviewed only TransFSW (MC-TransFSW) had a lower mean degree of 2.40. Finally, a very small fraction of clients reviewed only DK sex workers (0.3%).

We examined cross-country review activity by counting the number of countries or territories in which the sex workers reviewed by each client were listed. Clients were grouped according to this number (one, two, three, or four or more countries/territories), and the mean degree, defined as the number of distinct reviewed sex workers, was calculated for each group (Table 2). Across all client groups, clients whose reviews involved a larger number of countries or territories had higher mean degrees, indicating that broader cross-country review activity was associated with a larger number of distinct reviewed sex workers.

Notably, MC-CisTransFSW clients showed broader cross-country review activity than both MC-CisFSW and MC-TransFSW clients, and their mean degree increased sharply with the number of countries or territories represented in their reviews. In contrast, MC-TransFSW clients showed the narrowest geographic range and the lowest mean degree. These trends are further illustrated by a scatter plot of the number of countries or territories represented in each client's reviews and the number of reviews posted, which highlights the presence of highly active MC-CisFSW and MC-CisTransFSW clients linked to sex workers listed in multiple countries or territories (Supplementary Fig. S1).

Figure 4B shows the complementary cumulative degree distributions of clients classified as MC-CisFSW, MC-CisTransFSW, and MC-TransFSW. For MC-CisTransFSW clients, degrees were further separated according to whether the reviewed sex workers were CisFSW or TransFSW. Among reviews involving CisFSW, the distribution for MC-CisTransFSW clients lay above that for MC-CisFSW clients, indicating that MC-CisTransFSW clients tended to review a larger number of CisFSW than clients who reviewed only CisFSW. Similarly, among reviews involving TransFSW, the distribution for MC-CisTransFSW clients lay above that for MC-TransFSW clients, indicating that MC-CisTransFSW clients also tended to review a larger number of TransFSW than clients who reviewed only TransFSW. The complementary cumulative degree distributions of sex workers showed that CisFSW had a slightly higher degree distribution than TransFSW, reflecting a larger number of distinct reviewing clients per sex worker among CisFSW (Fig. 4C).

To evaluate whether the observed network structure differed from that expected under the degree sequence alone, we generated 100 degree-preserving randomized bipartite networks by rewiring edges while preserving the degree of each node. We then compared the observed bipartite clustering coefficients and degree correlations with those obtained from the randomized networks (Fig. 6A,B).

[Figure 6 near here]

The bipartite clustering coefficient of the full *TER*-derived network was 0.018, which was higher than those of the randomized networks (Fig. 6A). This indicates that four-cycle closure was more frequent than expected from the degree sequence alone, consistent with the presence of locally clustered client–sex worker groups. Elevated clustering relative to the null model was also observed across temporal subnetworks, with particularly high values in the 2018–2020 and 2021–2023 periods.

The observed degree correlation was generally more negative than those of the randomized networks, indicating disassortative mixing between clients and sex workers beyond that expected from the degree sequence alone (Fig. 6B and Supplementary Figs. S2–S14). However, this pattern was not uniform across all temporal windows. In particular, the 2018–2020 and 2021–2023 subnetworks showed higher degree correlation values than their randomized counterparts, whereas the 2009–2011 subnetwork showed only a small difference from the null distribution.

Finally, we examined average shortest path lengths on the giant connected component. The full aggregated network had an average path length of 6.10, which was shorter than those of the three-year temporal subnetworks, ranging from 6.19 to 8.05 (Fig. 6C). Thus, the long-term aggregated network was more compact than networks restricted to shorter time windows. Together with the elevated bipartite clustering, this result suggests that the *TER*-derived network exhibits small-world-like organization, shaped not only by cross-country review ties but also by temporal aggregation over long-term users.

## Discussion

This study constructed and analyzed a large bipartite commercial-sex review network derived from *TER* reviews. By extending previous analyses of online commercial-sex networks from national or metropolitan settings to a cross-country setting, our study characterizes how clients, sex worker profiles, locations, and time periods are connected within a single large-scale network spanning multiple countries and territories. The central finding is that the *TER*-derived network is highly connected despite its geographic breadth and strong concentration in the United States. The giant connected component contained most clients and sex worker profiles, and the average shortest path length of the full network was 6.10, shorter than those reported for several previously studied national networks (Table 1). Together with bipartite clustering that was higher than expected under degree-preserving randomization, these results indicate that the network exhibits small-world-like organization at the cross-country scale.

This compact structure appears to arise from two complementary forms of bridging. Spatially, a small fraction of clients had review ties spanning multiple countries or territories, thereby connecting sex worker profiles located in otherwise distant regions. This pattern extends previous findings from Japan, where clients moving across regions acted as weak ties connecting geographically separated parts of the commercial-sex network [19], suggesting that similar bridging can also occur at the cross-country scale. Temporally, long-term users linked different periods of the dataset, making the 25-year aggregated network more compact than networks constructed from shorter time windows. Thus, the small-world-like structure observed here is not simply a consequence of local clustering within cities or countries, but reflects the combined effects of cross-country review ties and long-term persistence in the network.

The temporal component of this compactness deserves particular attention. For example, 1.1% of sex worker profiles and 2.8% of clients who were active in 2000–2002 were still active in 2021–2023, thereby connecting early and recent parts of the network (Tables S3–S4). Among sex worker profiles that received more than two reviews, the mean interval between the first and last reviews was 17.34 months; among clients who posted more than two reviews, the corresponding interval was 32.65 months. These averages were influenced by a long tail of persistent users, since most individuals had much shorter observed activity intervals (Supplementary Fig. S15). Thus, the average path length of the 25-year aggregated network should not be interpreted as a time-respecting transmission path, but rather as evidence that long-term user activity contributes to the compactness and structural continuity of the TER-derived review network.

The observed bipartite clustering coefficient of the *TER*-derived network (0.018) was lower than that reported for the Japanese commercial-sex network (0.028) [19], but remained clearly higher than expected under degree-preserving randomization. This indicates that local closure among clients and sex workers persists even in a much larger and more geographically dispersed network. The lower clustering compared with the Japanese network may reflect differences in platform use, geographic scale, and temporal coverage: the *TER* dataset spans about 25 years and covers many countries, whereas the Japanese dataset was restricted to a shorter, national-scale observation window. Such broader spatial and temporal aggregation may dilute tightly connected local groups.

At the same time, the elevated clustering suggests that local organizing mechanisms remain important. Previous work showed that establishments and service locations can generate clustered contact patterns, with clients tending to review multiple sex workers associated with the same establishment or area [19]. Similar mechanisms, including geographic proximity and agency affiliation, may also shape clustering in *TER*, although these factors were not fully observable in our dataset. The higher clustering coefficients observed after 2020 may likewise reflect changes in travel and contact patterns during and after the COVID-19 pandemic, but this interpretation remains tentative.

The degree distributions of both clients and sex worker profiles were highly heterogeneous, with a small number of nodes having many review ties. We do not attempt to classify these distributions as scale-free or to identify a specific parametric form. Rather, the important point for the present analysis is that this heterogeneity concentrates connectivity in a small subset of highly active clients and sex worker profiles, thereby contributing to cross-country bridging and short path lengths in the aggregated network.

The degree distribution also differed across client groups defined by the *TER* profile categories of the sex worker profiles they reviewed. Clients who reviewed both CisFSW and TransFSW profiles (MC-CisTransFSW) had higher degrees than clients who reviewed only CisFSW or only TransFSW profiles. In particular, MC-CisTransFSW clients tended to review more CisFSW profiles than MC-CisFSW clients and more TransFSW profiles than MC-TransFSW clients, indicating that this group was not simply intermediate between the two single-category client groups but was instead composed of highly active reviewers. Among clients with two or more reviews, the mean degree was 13.6 for MC-CisTransFSW clients, compared with 8.7 for MC-CisFSW clients and 4.5 for MC-TransFSW clients.

MC-CisTransFSW clients also showed broader cross-country review activity, with the highest proportion of clients whose review ties spanned two or more countries or territories (Table 2). Thus, review ties involving TransFSW profiles were structurally concentrated around a relatively small group of highly active clients with broader geographic reach in the *TER*-derived review network. This pattern suggests that the position of TransFSW-related contacts in the broader network is shaped not only by the degree of TransFSW profiles themselves, but also by the activity and cross-country connectivity of the clients who reviewed them.

The *TER*-derived network also exhibited negative degree correlation between clients and sex worker profiles, consistent with previous studies of online commercial-sex networks [17–19]. This disassortative pattern indicates that high-degree sex worker profiles tend to be connected to many low-degree clients, while high-degree clients tend to review sex worker profiles with fewer reviews. Such a pattern may reflect heterogeneity in client experience, search behavior, and sex worker visibility, although the present data do not allow us to distinguish among these mechanisms. The strength of degree correlation also varied across temporal subnetworks, with some periods showing larger deviations from degree-preserving randomized networks than others (Fig. 6B). These temporal differences may reflect platform-specific disruptions or broader external shocks, but their mechanistic explanation remains a topic for future work.

Our constructed commercial-sex review network has implications for understanding contact structures relevant to STI transmission. Degree heterogeneity, shortest path lengths, and clustering all shape epidemic processes on sexual networks. Among these properties, the degree distributions in the *TER*-derived network revealed heterogeneity in contact patterns that may be relevant to STI transmission risk. Notably, clients who reviewed sex workers in multiple countries or territories had heavier-tailed degree distributions than those who reviewed them in a single country or territory. This pattern suggests that cross-country review activity is associated with highly heterogeneous contact opportunities and may create structural conditions under which infections could spread across regions.

While previous studies show international travelers had higher number of sexual partnerships and increased STI acquisition risk [33–35], our study provides a possible network-structural perspective on such associations. These findings would contribute to developing mathematical models to simulate international dynamics of STIs, as its demand has been increasing due to the concerns about the international disseminations of mpox, zika and antimicrobial strains of STIs (e.g. gonorrhoea and mycoplasma genitalium). Also, more targeted public health approaches, including pre-travel sexual health advice may be worth further investigation while the appropriate interventions for cross-country commercial-sex contacts remain an open question [36].

Our analysis of the *TER* revealed the structural concentration of review ties involving TransFSW profiles. Transgender women sex workers are facing several barriers to sexual health services and elevated HIV-related vulnerabilities in many settings [37,38]. However, quantitative data on the sexual contact patterns for those sex workers and their clients are quite limited. In the *TER*-derived network, even though the degree distribution for TransFSW and CisFSW are found to be similar, their client review patterns are quite distinct. For example, TransFSW received reviews from clients who reviewed both CisFSW and TransFSW, which accounted for 54.8% (Figure 5). However, these MC-CisTransFSW had disproportionately heavier tails than MC-CisFSW or MC-TransFSW. This suggests that at least with this sexual review network, the risk of the STI transmission may be difficult to capture using egocentric surveys alone, and the client sexual behavior as well as sex worker behavior should be collected for the precise assessment of STI transmission risks.

There are several limitations in our study. First, the TransFSW-related findings should be interpreted as a structural pattern in *TER* review ties rather than as a direct measure of gender identity, sexual preference, or individual risk behavior. The *TER* profile category used to identify TransFSW was a profile-based classification, and our data captured only reviewed commercial-sex encounters. We could not observe unreviewed contacts, non-commercial partnerships, condom use, drug use, or other behavioral and contextual factors known to shape STI risk [39,40]. Nevertheless, the concentration of TransFSW-related review ties around a small group of highly active clients suggests a form of structural exposure that may be relevant for future studies of sexual health and STI transmission in commercial-sex networks.

Secondly, the *TER*-derived network represents only a partial and platform-specific subset of commercial-sex contacts. It excludes many forms of transactional sex not represented on *TER*, and not all clients who use commercial-sex services post reviews. The proportion of encounters recorded as reviews is therefore unknown. In addition, usernames are pseudonymous online identifiers, and we cannot independently verify whether they correspond one-to-one to real individuals over time. The data also provide service locations but not the nationality, origin, or movement histories of sex workers. Thus, cross-country review ties should not be interpreted as complete records of client or sex worker mobility, and the network analyzed here should not be regarded as a complete representation of commercial-sex contacts in the United States or worldwide.

Despite these limitations, *TER* provides an unusually large, geographically diverse, and temporally extensive source of reviewed commercial-sex contacts. Its long-term archive enabled us to construct a 25-year bipartite network and to examine cross-country connectivity, temporal bridging, clustering, degree mixing, and contact concentration involving TransFSW within a single empirical framework. By analyzing this distinctive dataset, our study shows how large-scale connectivity can emerge in a commercial-sex review network through a small number of cross-country and long-term review ties, providing a structural basis for future network-based studies of STI transmission.

**List of abbreviations**

CisFSW: Cisgender female sex worker

DK: Don’t know

GCC: Giant connected component

MC: Male client

STI: Sexually transmitted infection

TER: The Erotic Review

TransFSW: Transgender female sex worker

**Ethics approval and consent to participate**

The Human Research Ethics Committee of Shizuoka University determined on 21 May 2024 that formal ethics approval and informed consent were not required for this study because it used de-identified, publicly visible online data and did not involve direct participation, intervention, interaction with individuals, or collection of identifiable private information.

**Availability of data and materials**

The individual-level review and network data analyzed in this study are not publicly available because they contain pseudonymous identifiers and sensitive information about reported commercial-sex contacts. Aggregated data supporting the main findings are included in this article and its Additional file 1. Additional aggregated data and de-identified analytical datasets may be made available from the corresponding author upon reasonable request, subject to ethical and privacy considerations.

**Competing interests**

The authors declare that they have no competing interests.

**Funding**

This work was supported by the Japan Society for the Promotion of Science (JSPS) KAKENHI Grant Numbers JP25K01465 (HI and SM), JP25K22145 (HI and SM), JP25K07150 (SM and HI), JP23KK0210 (HI), and the Joint Usage / Research Center on Tropical Disease, Institute of Tropical Medicine, Nagasaki University (2025-Seeds-02). The funders had no role in the study design, data collection, analysis, interpretation of data, writing of the manuscript, or decision to submit the article for publication.

**Authors' contributions**

HI and SM conceived the study. KS collected the raw data. SM curated and preprocessed the data, developed the analysis code, and conducted the network analyses. HI and SM designed the analytical framework, interpreted the results, and prepared the figures and tables. TRA contributed to the public health and STI-related framing, interpretation, and manuscript revision. HI and SM drafted the manuscript. All authors reviewed, edited, and approved the final manuscript.

**Acknowledgments**

The authors are grateful to Dr. Elizabeth Fearon for her valuable comments and feedback from the perspectives of public health and STI epidemiology on this paper. The authors used ChatGPT (OpenAI) to assist with improving the clarity and fluency of the English text.

## Figures and Tables

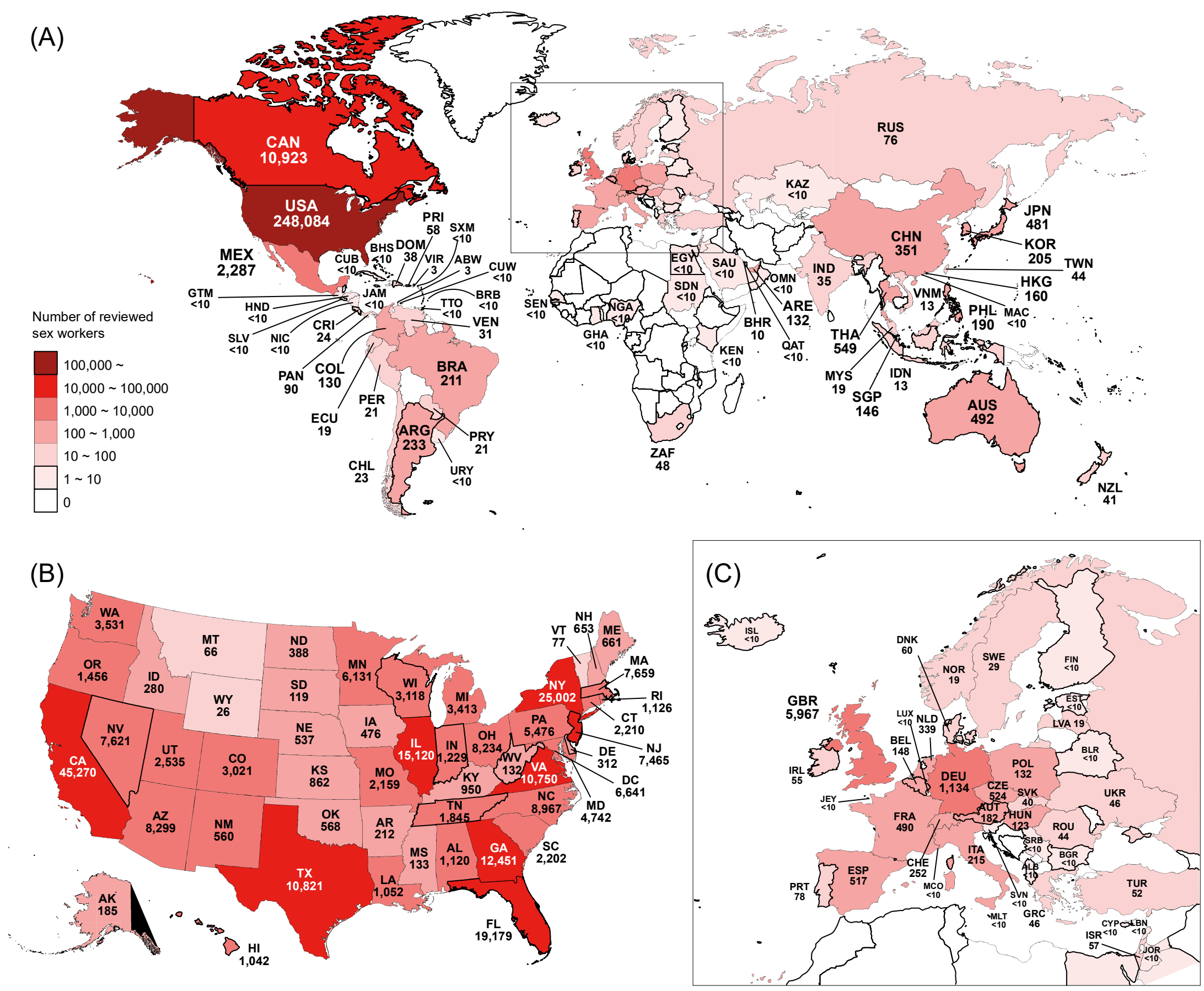

**Figure 1. Number of reviewed sex workers in each region.** Number of reviewed sex workers in each region. The spatial distribution of reviewed sex worker counts, with dark red representing higher numbers, is visualized (**A**) globally, (**B**) for the U.S. states, and (**C**) in Europe. For the 26 sex workers who provided services in multiple states within the U.S., we counted them based on the state where they received their first review. One sex worker was excluded from this figure because of their choice of 'Other' as their working place. The three-letter abbreviations correspond to alpha-3 ISO country codes, and the two-letter abbreviations correspond to U.S. state and territory abbreviations. The maps are modified from the World Map (world-09) and the U.S. Map (us-02) by Vemaps.com.

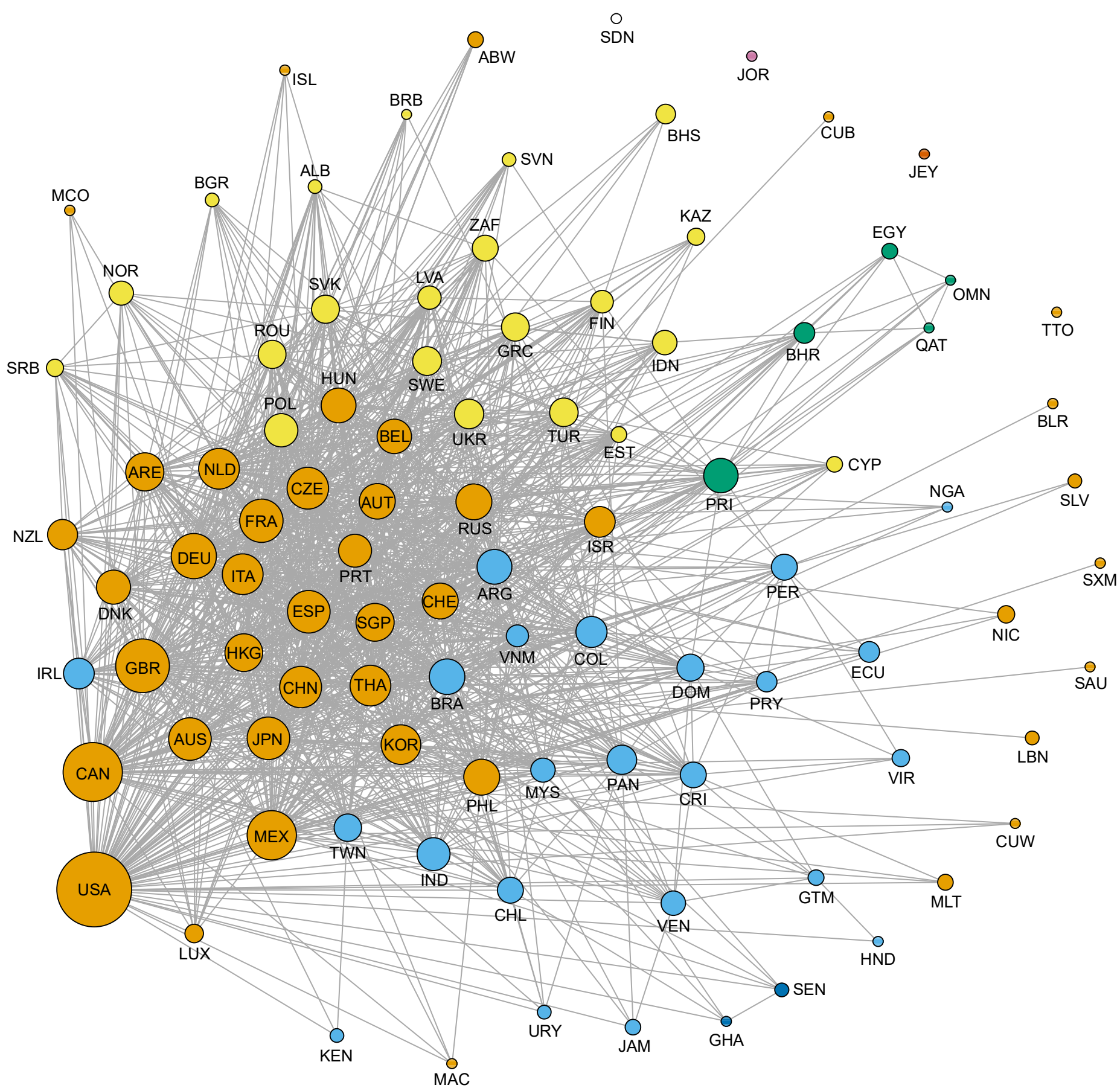

**Figure 2. Country-level projection of cross-country review ties.** The circles represent the countries/territories where sex workers provided services, and their sizes are proportional to the number of reviews (shown on a logarithmic scale). A link between two countries or territories indicates that at least one client posted reviews for sex workers listed in both locations. The three-letter abbreviations correspond to alpha-3 ISO country codes. The network is visualized using the Fruchterman-Reingold algorithm. Node colors indicate communities detected by the Louvain method and are used only to aid visual interpretation; these community assignments were not used in subsequent analyses.

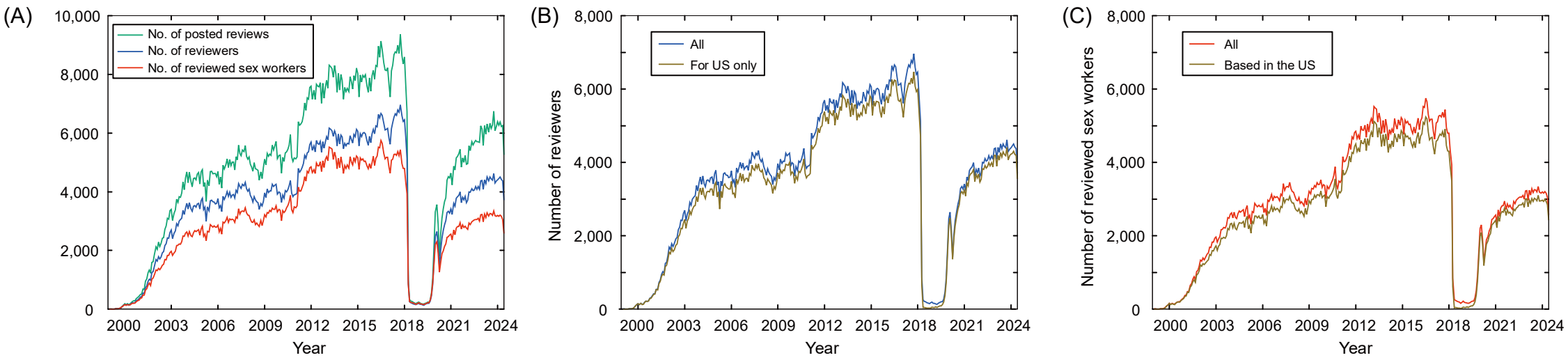

**Figure 3. Temporal patterns of review activity in the TER-derived dataset.** (**A**) Monthly total numbers of reviews posted, reviewers, and reviewed sex workers. (**B**) Monthly total numbers of reviewers and those who posted reviews exclusively for sex workers providing services in the U.S. (**C**) Monthly total numbers of reviewed sex workers and those providing services in the U.S.

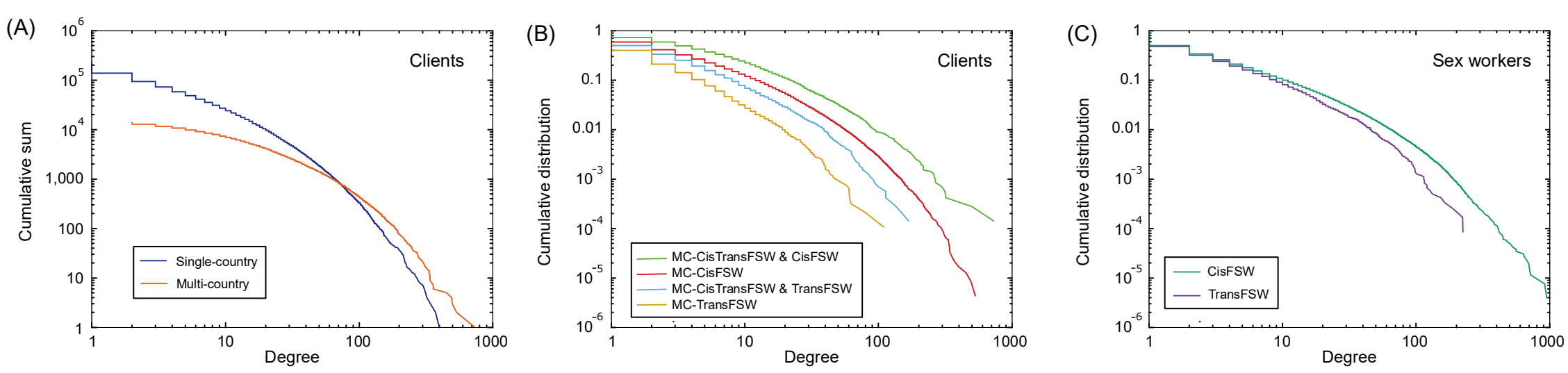

**Figure 4. Degree distributions by client and sex worker category. (A)** Complementary cumulative degree distributions of clients whose review ties were limited to one country or territory and those whose review ties spanned multiple countries or territories. Degree is the number of distinct sex workers reviewed. The vertical axis shows the number of clients with degree greater than or equal to the corresponding value. **(B)** Complementary cumulative degree distributions of client groups defined by the TER profile categories of reviewed sex workers. For MC-CisTransFSW clients, degrees are separated into review ties to CisFSW and TransFSW. **(C)** Complementary cumulative degree distributions of CisFSW and TransFSW, where degree is the number of distinct reviewing clients. In (B) and (C), the vertical axis shows the proportion of nodes with degree greater than or equal to the corresponding value.

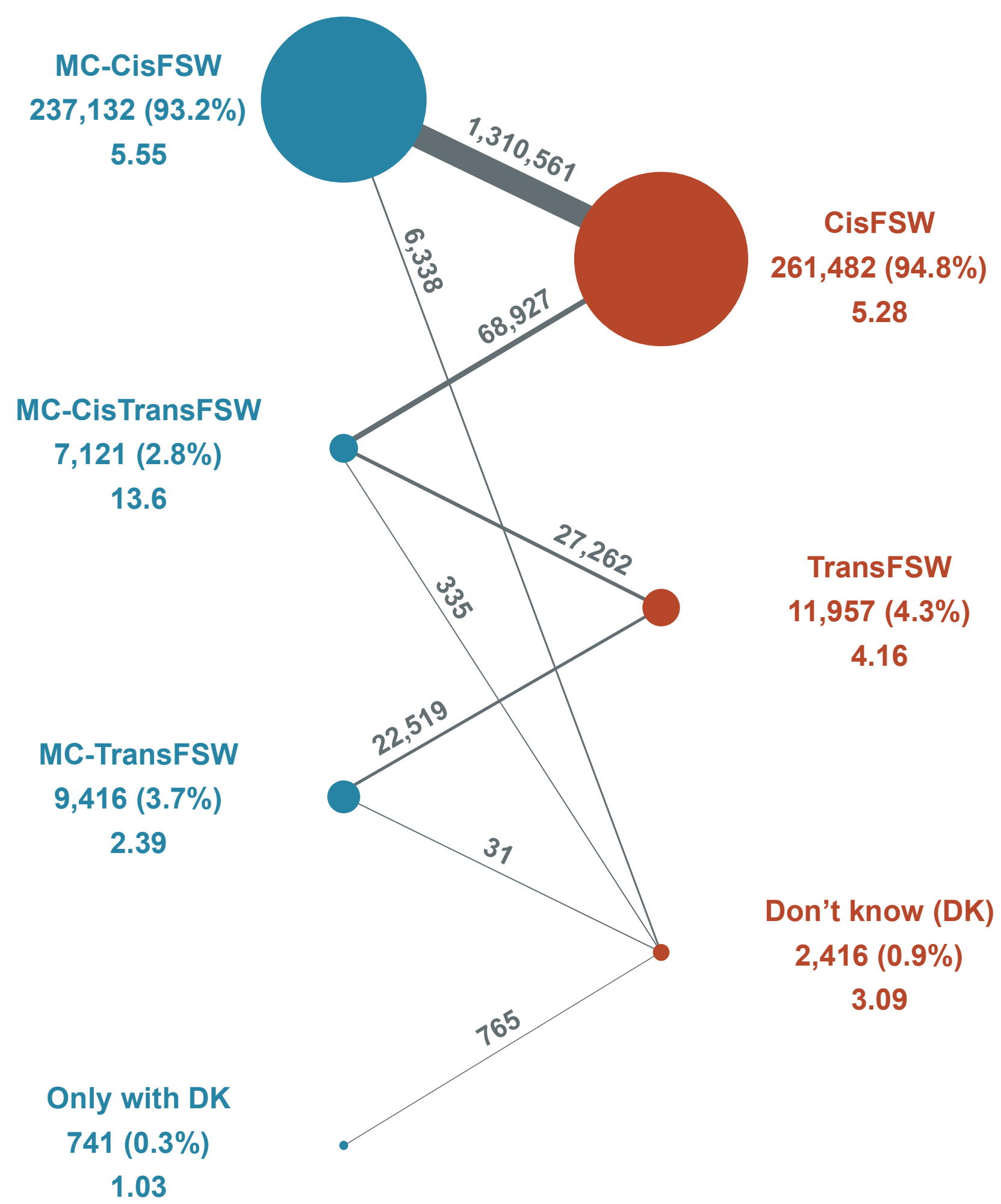

**Figure 5. Relationships between the four client types and the three sex worker types.** Each circle represents a type of client (blue) or sex worker (red) and its size is proportional to the square root of the number of individuals. The thickness of the links between each client type and sex worker type is proportional to the square root of the number of reviews (grey numbers). The following information is displayed beside the circles: i) the type name, ii) the number of individuals and the proportion each type represents among all reviewers or all sex workers, and iii) the average degree.

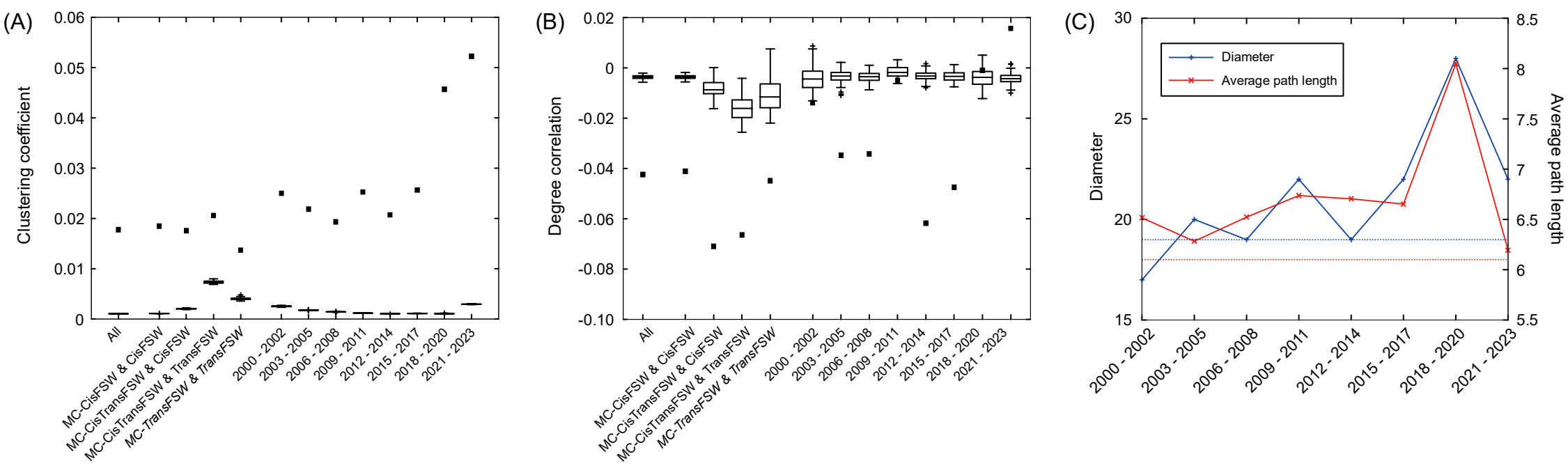

**Figure 6. Bipartite clustering, degree correlation, and path lengths. (A)** Bipartite clustering coefficients and **(B)** degree correlations calculated from the observed networks and degree-preserving randomized bipartite networks. The network constructed from the full TER dataset is denoted as “ALL.” The network was further divided into four contact-type subnetworks and eight temporal subnetworks based on review posting periods. Black dots indicate values calculated from the observed networks, and boxplots show the distributions obtained from 100 degree-preserving randomized networks. **(C)** Diameter and average shortest path length calculated for three-year temporal subnetworks. Dotted lines indicate the diameter (19) and average shortest path length (6.10) of the full aggregated network.

**Table 1. Characteristics of sexual contact networks of The Erotic Review (TER) in this study and other review sites in the sex industry.**

| | TER | Japan[b] | UK[c] | Brazil[d] |
|---|---|---|---|---|
| Study period for each dataset | Oct. 1999–Jun. 2024 | Nov. 2018–Apr. 2021 | Jan. 2014–Dec. 2017 | Sep. 2002–Oct. 2008 |
| Total number of clients | 254410 | 55824 | 2607 | 10109 |
| Total number of reviewed sex workers | 275855 | 17341 | 3870 | 6624 |
| Number of links excluding duplicates (number including duplicates) | 1436738 (1468852) | 89543 (121988) | 6426 | 40895 (50185) |
| Average number of reviewed sex workers per client | 5.65 | 1.60 | 2.46 | 4.05 |
| Average number of reviewers per sex worker | 5.21 | 5.16 | 1.66 | 6.17 |
| Size of GCC[a] | 509165 (96%) | 62917 (86%) | 3807 (59%) | 15810 (94.5%) |
| Second largest connected component | 12 | 18 | 125 | N.A. |
| Average path length in GCC | 6.10 | 9.87 | 9 | 5.78 |
| Diameter in GCC | 19 | 33 | 29 | 17 |
| Bipartite clustering coefficient | 0.018 | 0.028 | 0.12 | N.A. |
| Assortativity coefficient | -0.048 | -0.181 | -0.12 | -0.110 |

[a] GCC corresponds to the giant connected component.

[b,c,d] are from Ito et al. [19], Giommoni et al. [18], Rocha et al. [17], respectively.

**Table 2. Distribution of the number of countries each client posted reviews for.**

| Metrics | No. of countries each client reviewed for | Client type | | | |
|---|---|---|---|---|---|
| | | MC-CisFSW[a] | MC-CisTransFSW[b] | MC-TransFSW[c] | Only with DK |
| No. of clients | 1 | 224059 (94.49%) | 6254 (87.82%) | 9159 (97.27%) | 740 (99.87%) |
| | 2 | 10893 (4.59%) | 709 (9.96%) | 232 (2.46%) | 1 (0.013%) |
| | 3 | 1495 (0.63%) | 109 (1.53%) | 21 (0.22%) | 0 (0.00%) |
| | 4 or more | 685 (0.29%) | 49 (0.69%) | 4 (0.04%) | 0 (0.00%) |
| No. of reviews | 1 | 1045649 (79.40%) | 70476 (73.01%) | 20269 (89.88%) | 763 (99.74%) |
| | 2 | 186508 (14.16%) | 16770 (17.37%) | 1706 (7.57%) | 2 (0.26%) |
| | 3 | 48900 (3.71%) | 5402 (5.60%) | 434 (1.92%) | 0 (0.00%) |
| | 4 or more | 35841 (2.72%) | 3876 (4.02%) | 141 (0.63%) | 0 (0.00%) |
| Average degree | 1 | 4.7 | 11.3 | 2.2 | 1.0 |
| | 2 | 17.1 | 23.7 | 7.4 | 2 |
| | 3 | 32.7 | 49.6 | 20.7 | - |
| | 4 or more | 52.3 | 79.1 | 35.3 | - |

[a] MC-CisFSW corresponds to male clients who have sex with cisgender female sex workers.

[b] MC-CisTransFSW corresponds to male clients who have sex with cisgender and transgender female sex workers.

[c] MC-TransFSW corresponds to male clients who have sex with transgender female sex workers.

# Additional file 1: Supplementary figures and tables

## Cross-country structure of a sexual contact network derived from a commercial-sex review platform

Hiromu Ito, Keiko Shigeta, Toshiaki R. Asakura, Satoru Morita

Hiromu Ito, Satoru Morita

Email: ito.hiromu@nagasaki-u.ac.jp, morita.satoru@shizuoka.ac.jp

**Additional file includes:**

Figures S1 to S15

Tables S1 to S4

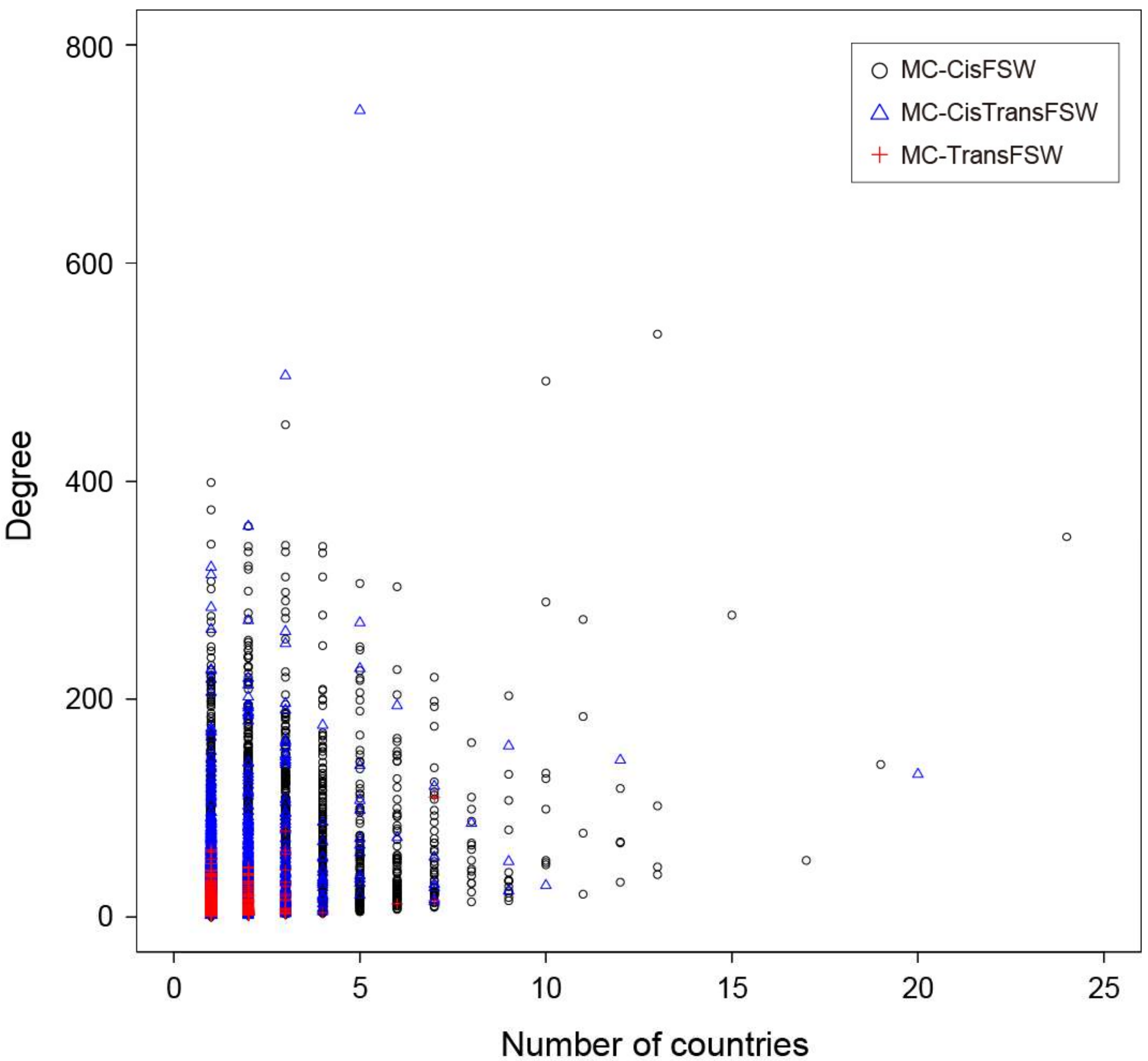

**Supplementary Fig. S1. Scatter plot of the number of countries or territories represented in each client's review ties versus client degree**. Each client type is represented by a different marker: MC-CisFSW (black circles), MC-CisTransFSW (blue triangles), and MC-TransFSW (red crosses).

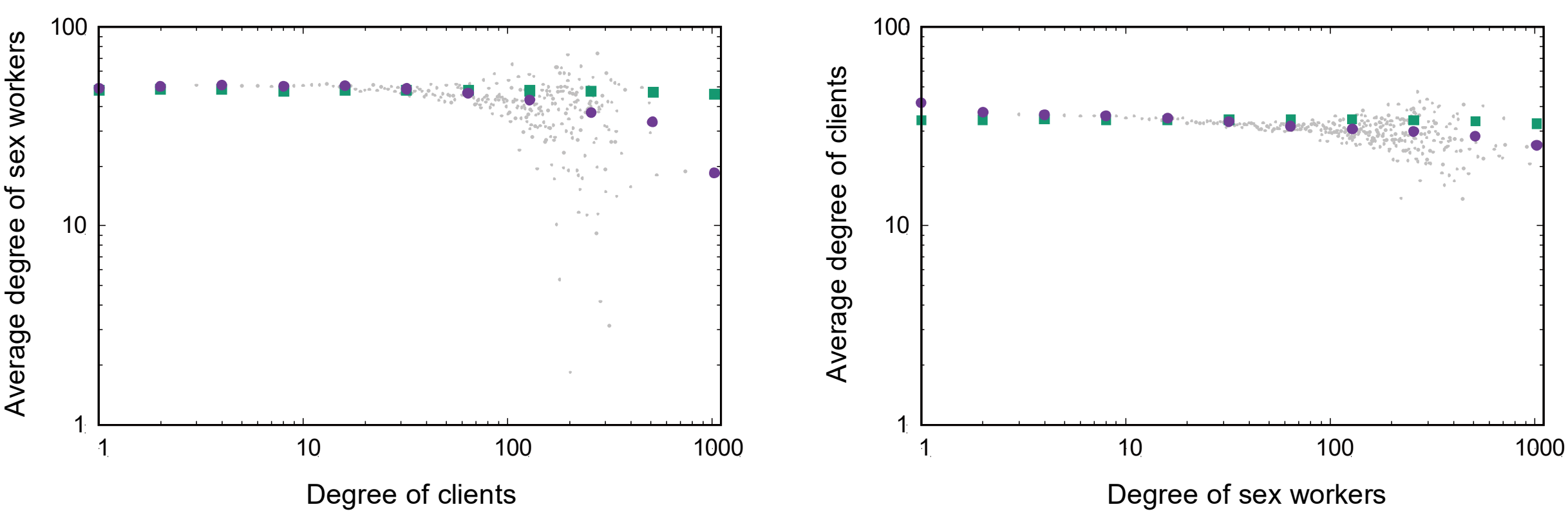

**Supplementary Fig. S2. Degree-correlation curves for the TER-derived client–sex worker review network.** The plots show the average nearest-neighbor degree of nodes with degree *k*. The left panel shows the degree of each client versus the average degree of neighboring sex workers, and the right panel shows the degree of each sex worker versus the average degree of neighboring clients. Gray points indicate raw node-level values in the observed network. Purple circles show logarithmically binned averages for the observed network, and green squares show logarithmically binned averages for degree-preserving randomized networks.

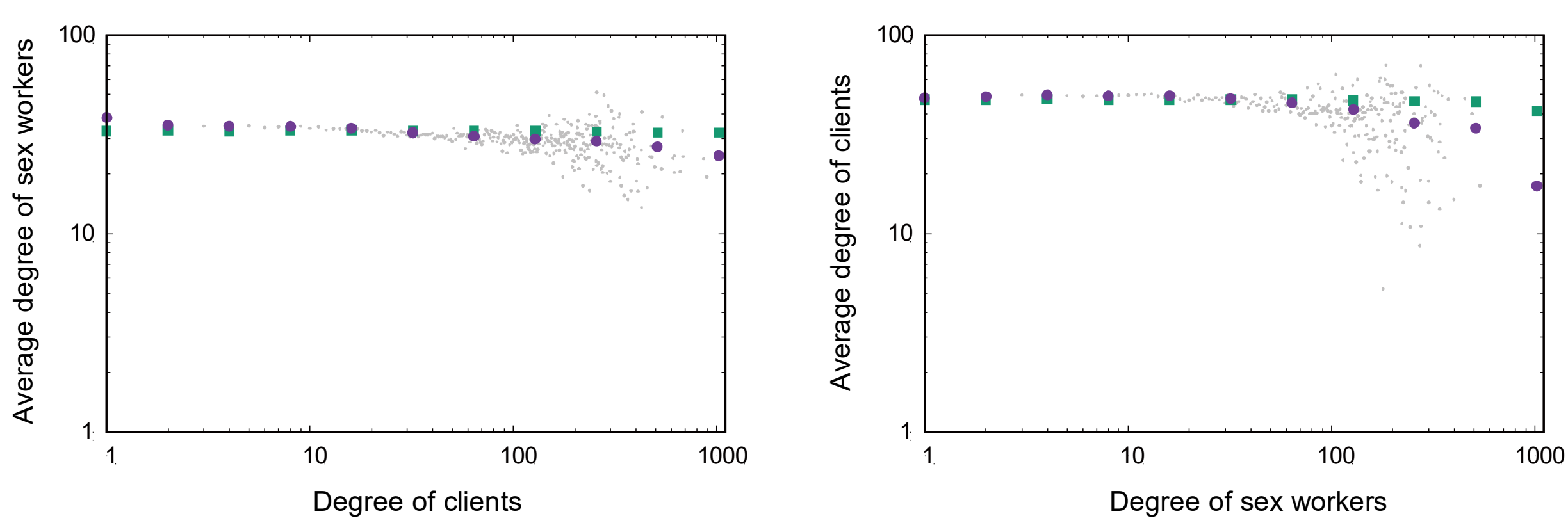

**Supplementary Fig. S3. Degree-correlation curves for review ties between MC-CisFSW and CisFSW.** The format is the same as in Supplementary Fig. S2.

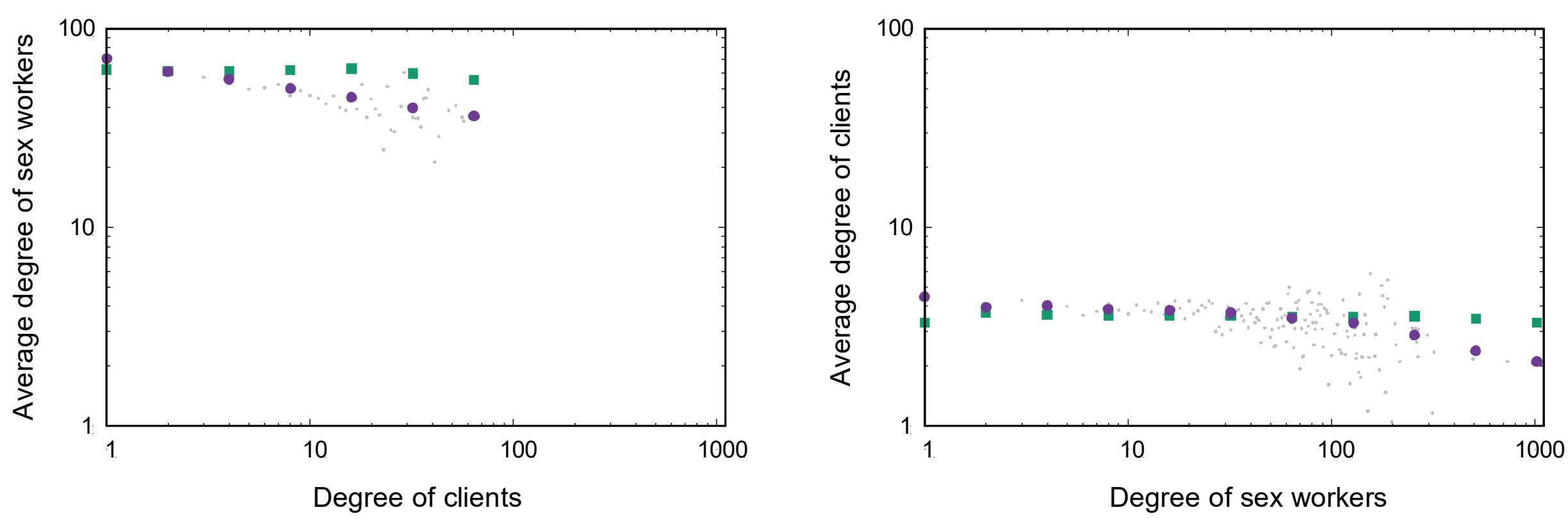

**Supplementary Fig. S4. Degree-correlation curves for review ties between MC-CisTransFSW and CisFSW.** The format is the same as in Supplementary Fig. S2.

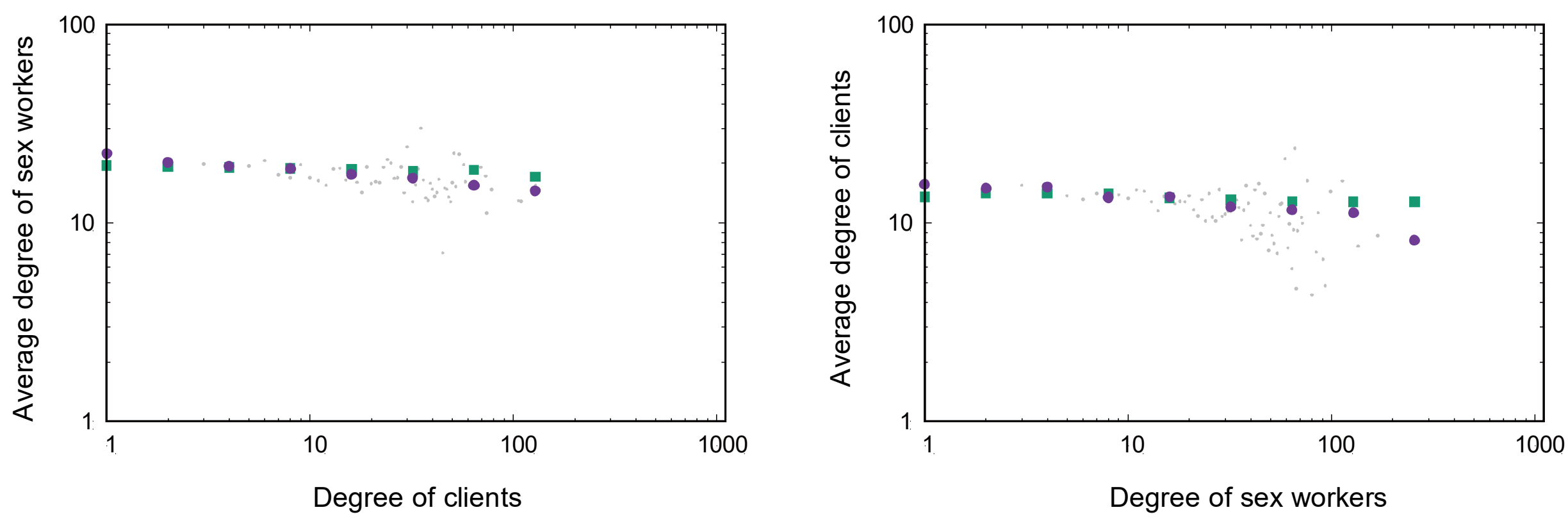

**Supplementary Fig. S5. Degree-correlation curves for review ties between MC-CisTransFSW and TransFSW.** The format is the same as in Supplementary Fig. S2.

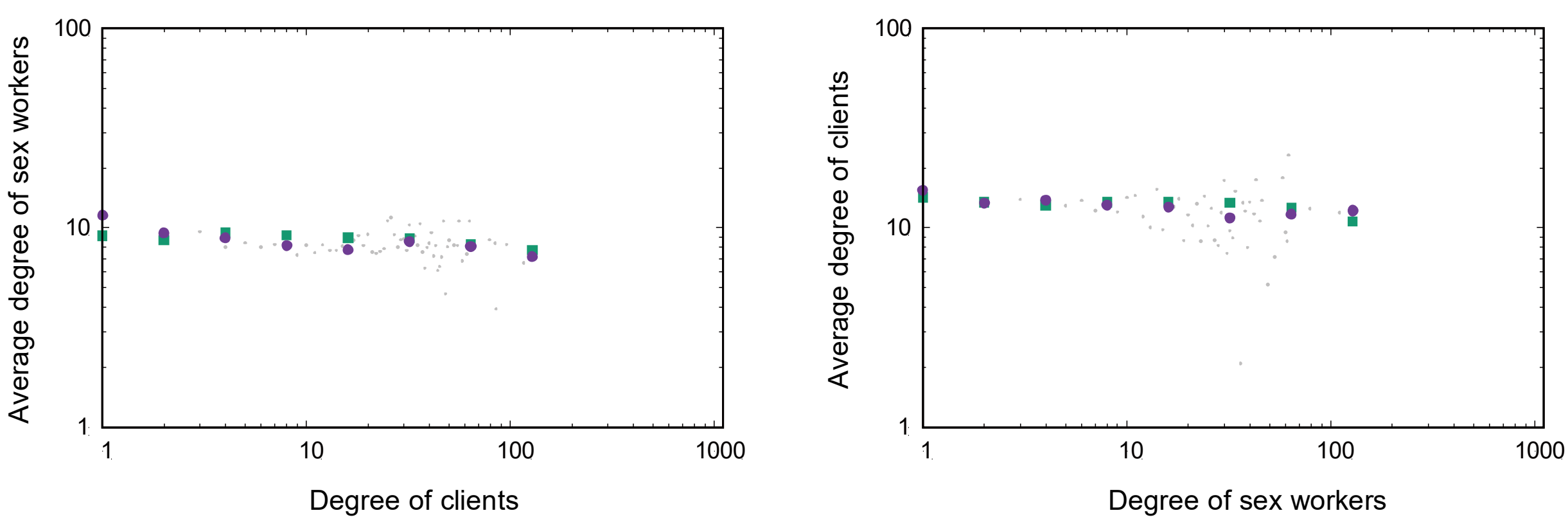

**Supplementary Fig. S6. Degree-correlation curves for review ties between MC-TransFSW and TransFSW.** The format is the same as in Supplementary Fig. S2.

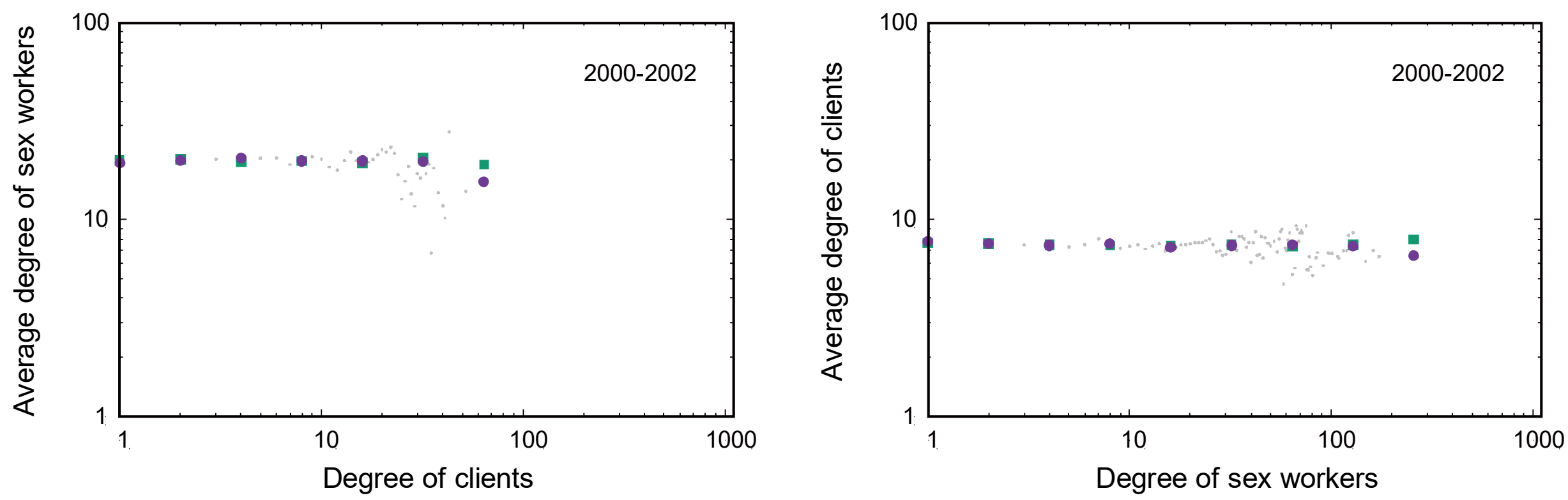

**Supplementary Fig. S7. Degree-correlation curves for the TER-derived client–sex worker review network during 2000–2002.** The format is the same as in Supplementary Fig. S2.

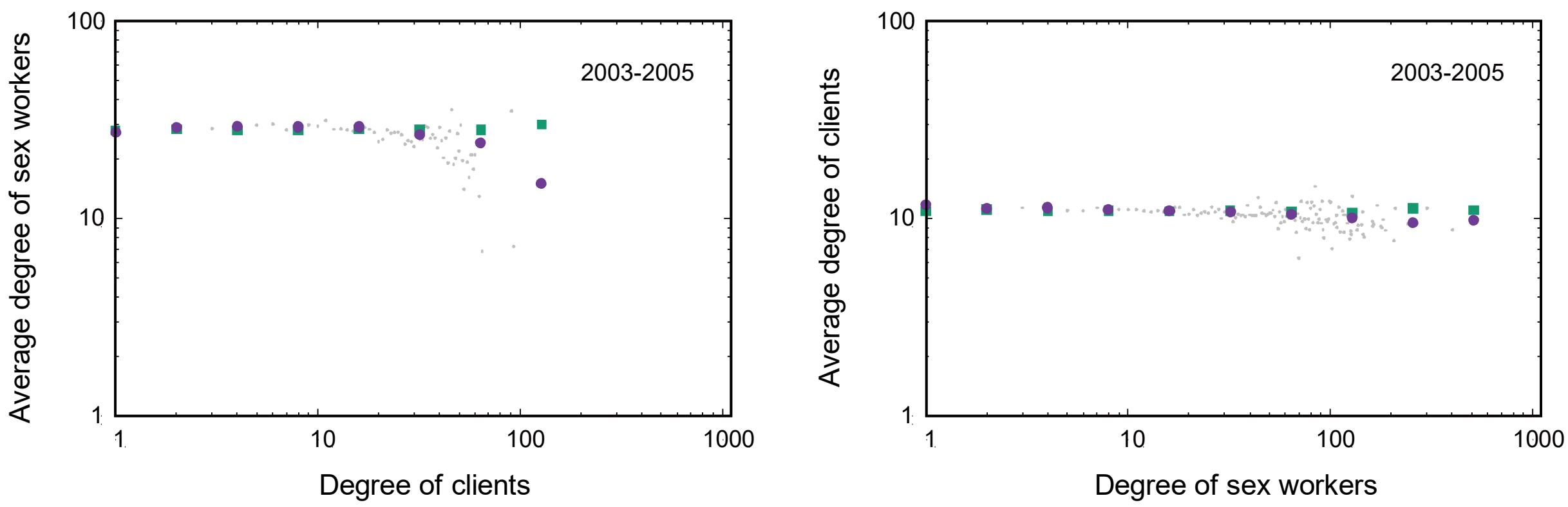

**Supplementary Fig. S8. Degree-correlation curves for the TER-derived client–sex worker review network during 2003–2005.** The format is the same as in Supplementary Fig. S2.

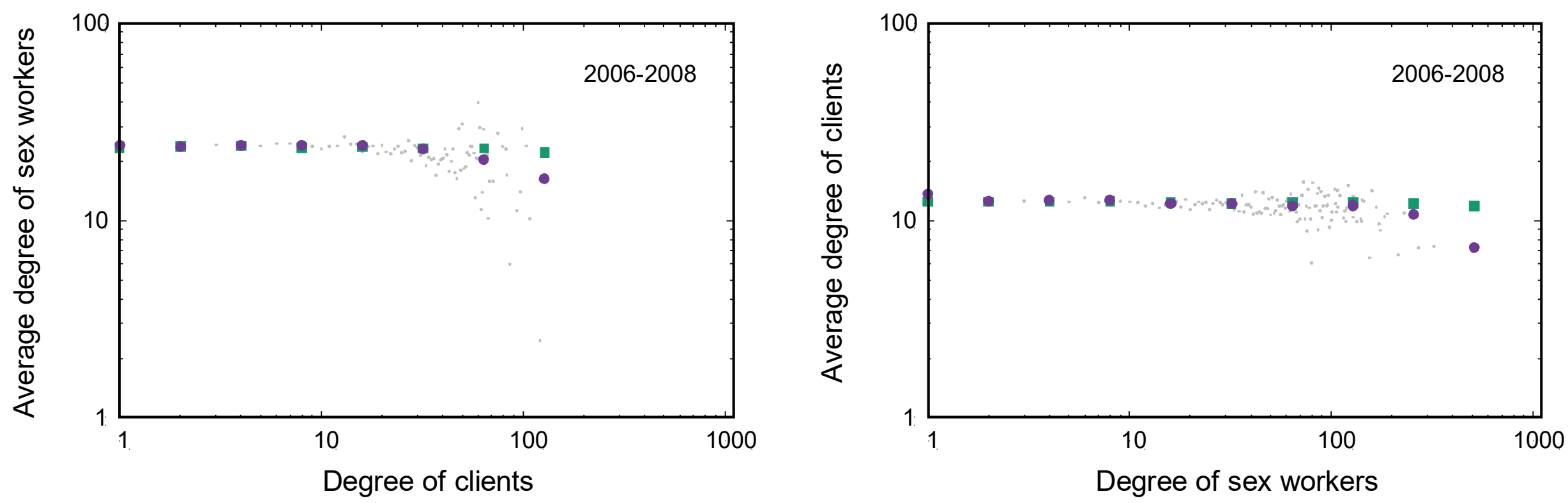

**Supplementary Fig. S9. Degree-correlation curves for the TER-derived client–sex worker review network during 2006–2008.** The format is the same as in Supplementary Fig. S2.

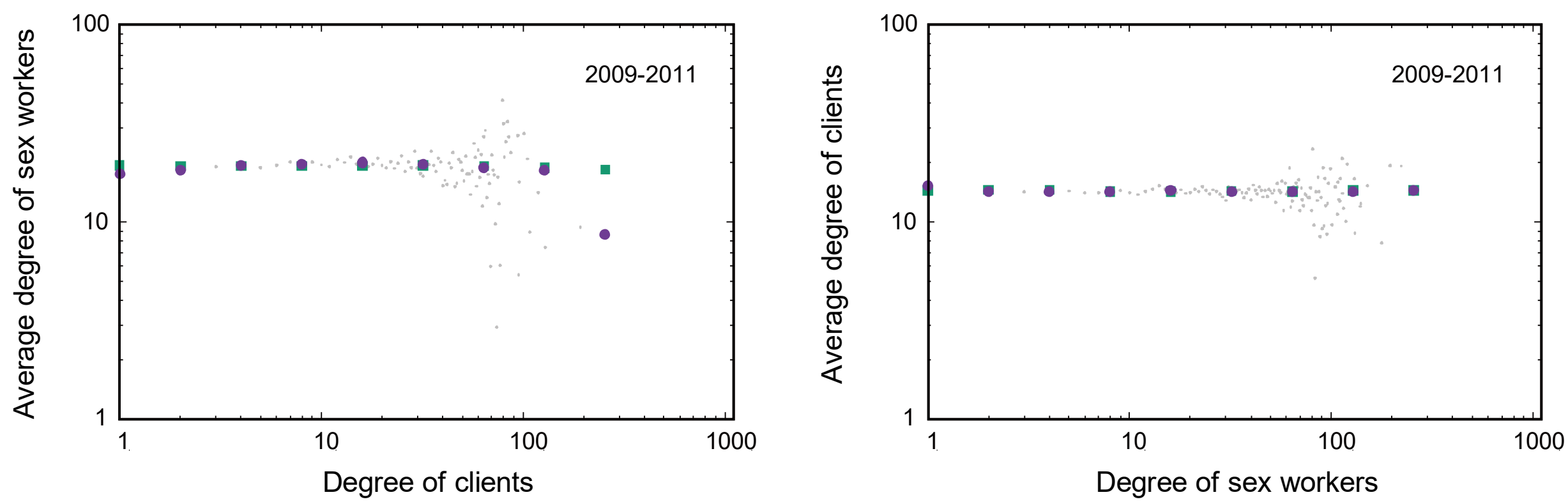

**Supplementary Fig. S10. Degree-correlation curves for the TER-derived client–sex worker review network during 2009–2011.** The format is the same as in Supplementary Fig. S2.

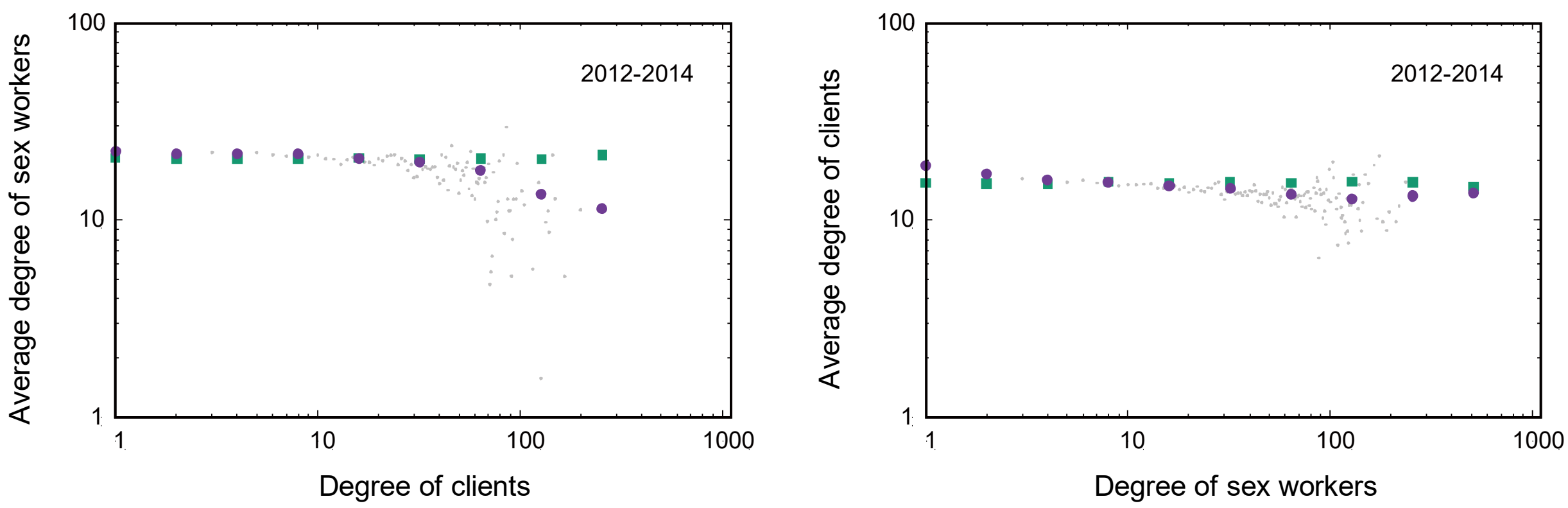

**Supplementary Fig. S11. Degree-correlation curves for the TER-derived client–sex worker review network during 2012–2014.** The format is the same as in Supplementary Fig. S2.

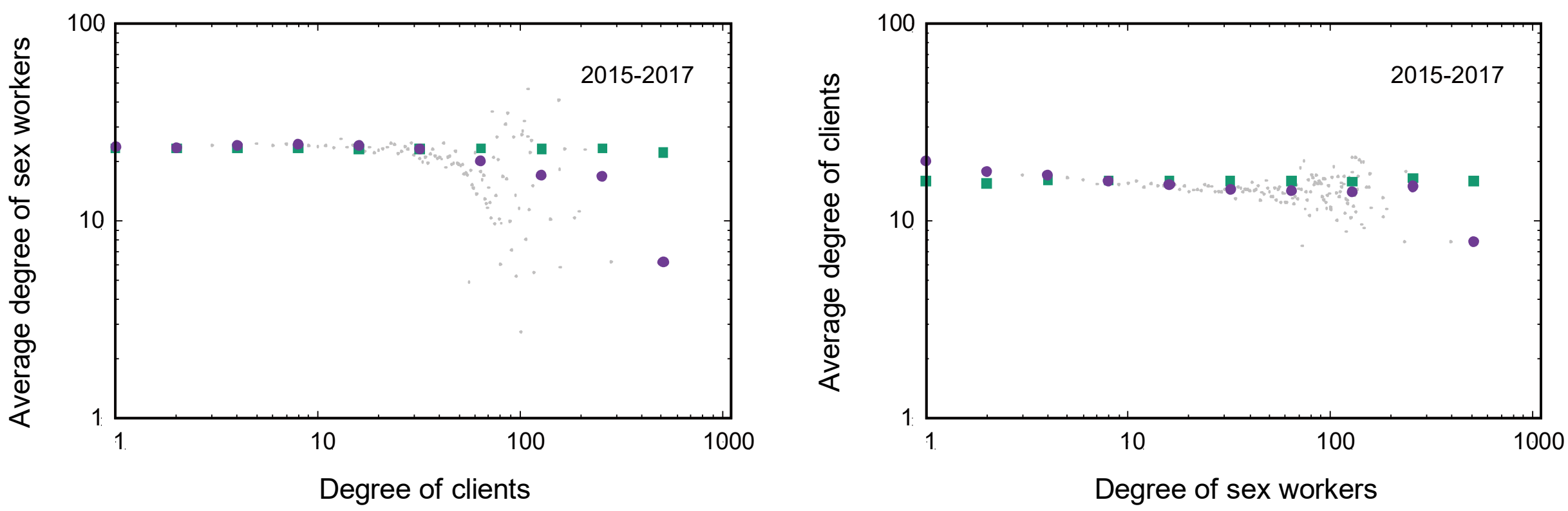

**Supplementary Fig. S12. Degree-correlation curves for the TER-derived client–sex worker review network during 2015–2017.** The format is the same as in Supplementary Fig. S2.

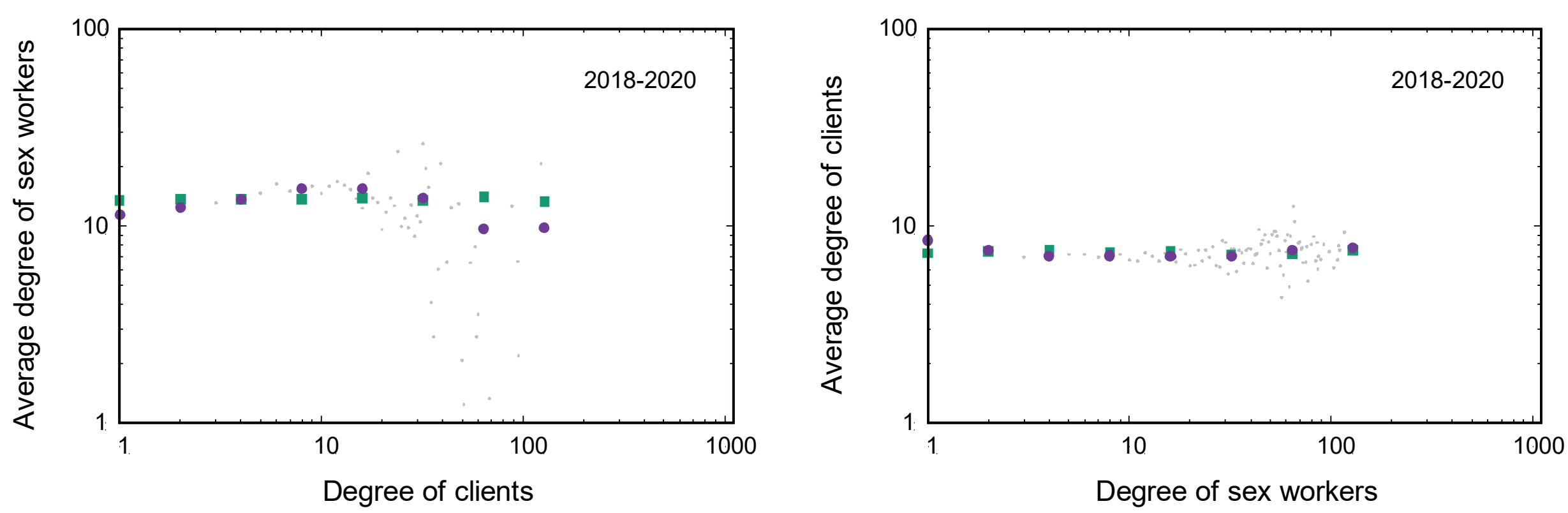

**Supplementary Fig. S13. Degree-correlation curves for the TER-derived client–sex worker review network during 2018–2020.** The format is the same as in Supplementary Fig. S2.

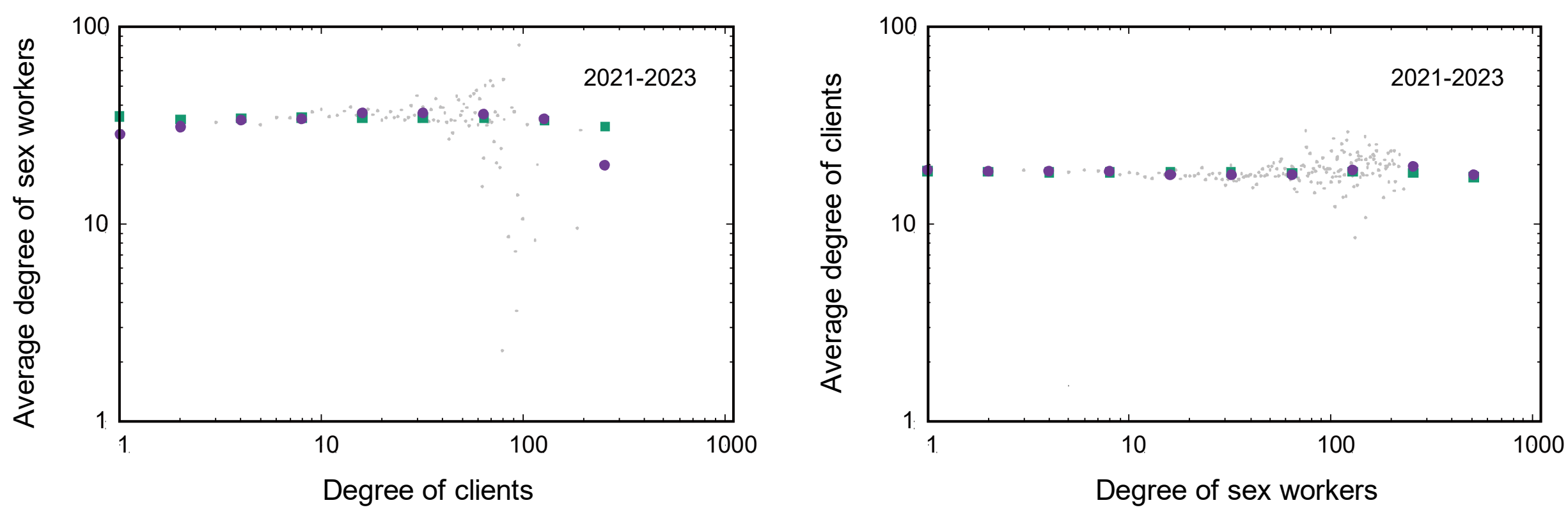

**Supplementary Fig. S14. Degree-correlation curves for the TER-derived client–sex worker review network during 2021–2023.** The format is the same as in Supplementary Fig. S2.

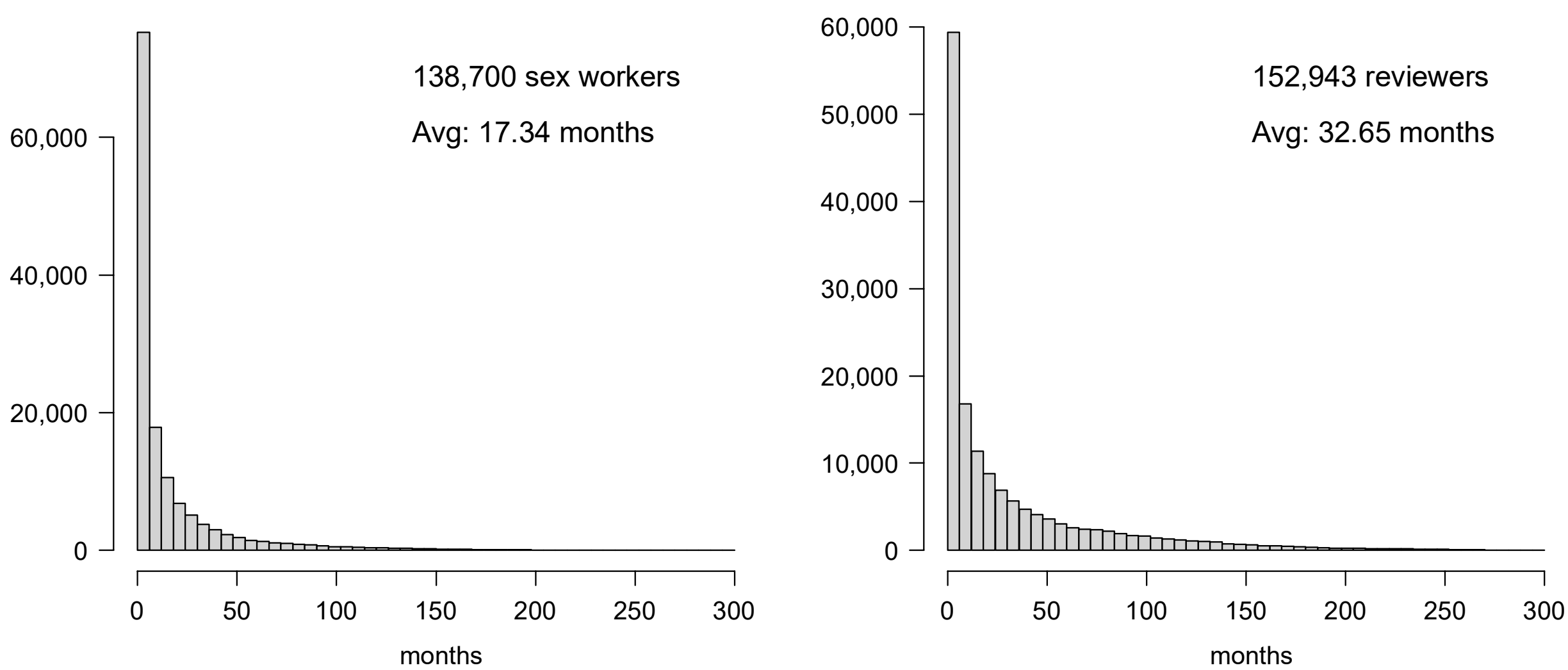

**Supplementary Fig. S15. Histogram of the time interval between the first and last reviews.** (**Left**) Histogram of the time interval between the first and last reviews for sex workers who received reviews more than twice. (**Right**) Histogram of the time interval between the first and last reviews for clients who posted reviews more than twice.

**Table S1. Network characteristics of each sex worker type**

| | CisFSW[a] | TransFSW[b] | | | Don't know | Total |
|---|---|---|---|---|---|---|
| | | Total | preop | postop | | |
| No. of sex workers | 261,482 (94.8%) | 11,957 (4.3%) | 11,711 (4.2%) | 246 (0.1%) | 2,416 (0.9%) | 275,855 (100%) |
| No. of received reviews | 1,379,488 (96.0%) | 49,781 (3.5%) | 48,749 (3.4%) | 1,032 (0.1%) | 7,469 (0.5%) | 1,436,738 (100%) |
| Average degree | 5.28 | 4.16 | 4.16 | 4.20 | 3.09 | 5.21 |

[a] CisFSW corresponds to cisgender female sex workers.

[b] TransFSW corresponds to transgender female sex workers.

**Table S2. Network characteristics of each client type.**

| | MC-CisFSW[a] | MC-CisTransFSW[b] | MC-TransFSW[c] | Only with DK | Total |
|---|---|---|---|---|---|
| No. of clients (reviewers) | 237,132<br>(93.2%) | 7,121<br>(2.8%) | 9,416<br>(3.7%) | 741<br>(0.3%) | 254,410<br>(100%) |
| No. of posted reviews | 1,316,866<br>(91.7%) | 96,524<br>(6.7%) | 22,550<br>(1.6%) | 765<br>(0.1%) | 1,436,738<br>(100%) |
| No. of reviews to | | | | | |
| CisFSW[d] | 1,310,561<br>(95.0%) | 68,927<br>(5.0%) | - | - | 1,379,488<br>(100%) |
| TransFSW[e] | - | 27,262<br>(54.8%) | 22,519<br>(45.2%) | - | 49,781<br>(100%) |
| Don't know (DK) | 6,338<br>(84.9%) | 335<br>(4.5%) | 31<br>(0.4%) | 765<br>(10.2%) | 7,469<br>(100%) |
| Average degree | 5.6 | 13.6 | 2.4 | 1.03 | 5.65 |
| Average degree (Reviewers with a degree of at least 2) | 8.7 | 13.6 | 4.5 | | |

[a] MC-CisFSW denotes male clients who reviewed CisFSW but not TransFSW.

[b] MC-CisTransFSW denotes male clients who reviewed both CisFSW and TransFSW.

[c] MC-TransFSW denotes male clients who reviewed TransFSW but not CisFSW.

[d] CisFSW corresponds to cisgender female sex workers.

[e] TransFSW corresponds to transgender female sex workers.

**Table S3. The number of sex workers who received reviews in both corresponding periods.** Diagonal numbers represent the number of sex workers who received reviews in this three-year period. The percentages represent the proportion of sex workers who received reviews spanning the periods indicated by both the rows and columns.

| Sex workers | 2000-2002 | 2003-2005 | 2006-2008 | 2009-2011 | 2012-2014 | 2015-2017 | 2018-2020 | 2021-2023 |
|---|---|---|---|---|---|---|---|---|
| 2000-2002 | 11,453 (100%) | 3,533 (30.8%) | 1,086 (9.5%) | 530 (4.6%) | 339 (3.0%) | 249 (2.2%) | 123 (1.1%) | 129 (1.1%) |
| 2003-2005 | | 31,046 (100%) | 5,841 (18.8%) | 1,857 (6.0%) | 1,104 (3.6%) | 706 (2.3%) | 337 (1.1%) | 328 (1.1%) |
| 2006-2008 | | | 41,848 (100%) | 7,034 (16.8%) | 2,793 (6.7%) | 1,631 (3.9%) | 729 (1.7%) | 621 (1.5%) |
| 2009-2011 | | | | 52,263 (100%) | 10,200 (19.5%) | 4,056 (7.8%) | 1,515 (2.9%) | 1,286 (2.5%) |
| 2012-2014 | | | | | 69,012 (100%) | 12,698 (18.4%) | 3,257 (4.7%) | 2,527 (3.7%) |
| 2015-2017 | | | | | | 66,672 (100%) | 8,419 (12.6%) | 4,858 (7.3%) |
| 2018-2020 | | | | | | | 21,428 (100%) | 7,174 (33.5%) |
| 2021-2023 | | | | | | | | 35,661 (100%) |

**Table S4. The number of clients who posted reviews in both corresponding periods.**

Diagonal numbers represent the number of clients who posted reviews in this three-year period. The percentages represent the proportion of clients who posted reviews spanning the periods indicated by both the rows and columns.

| Clients | 2000-2002 | 2003-2005 | 2006-2008 | 2009-2011 | 2012-2014 | 2015-2017 | 2018-2020 | 2021-2023 |
|---|---|---|---|---|---|---|---|---|
| 2000-2002 | 15,479 (100%) | 6,456 (41.7%) | 2,957 (19.1%) | 1,980 (12.8%) | 1,382 (8.9%) | 982 (6.3%) | 361 (2.3%) | 427 (2.8%) |
| 2003-2005 | | 41,614 (100%) | 12,359 (29.7%) | 6,698 (16.1%) | 4,424 (10.6%) | 3,107 (7.5%) | 1,115 (2.7%) | 1,317 (3.2%) |
| 2006-2008 | | | 49,394 (100%) | 14,868 (30.1%) | 8,525 (17.3%) | 5,828 (11.8%) | 1,965 (4.0%) | 2,254 (4.6%) |
| 2009-2011 | | | | 52,204 (100%) | 18,662 (35.7%) | 10,585 (20.3%) | 3,256 (6.2%) | 3,601 (6.9%) |
| 2012-2014 | | | | | 64,560 (100%) | 21,858 (33.9%) | 6,002 (9.3%) | 5,870 (9.1%) |
| 2015-2017 | | | | | | 70,814 (100%) | 12,728 (18.0%) | 9,352 (13.2%) |
| 2018-2020 | | | | | | | 26,051 (100%) | 10,251 (39.3%) |
| 2021-2023 | | | | | | | | 41,272 (100%) |